# Ghost Framing Theory:
# Exploring the role of generative AI in new venture rhetorical legitimation

Greg Nyilasy, PhD
Associate Professor
Department of Management and Marketing,
Faculty of Business and Economics,
University of Melbourne,
198 Berkeley St,
Carlton, VIC 3053,
Australia
gnyilasy@unimelb.edu.au

# Ghost Framing Theory: Exploring the role of generative AI in new venture rhetorical legitimation

## Abstract

Responding to the surging but largely invisible use of generative AI in entrepreneurial framing, I advance Ghost Framing Theory (GFT) to explain how hybrid founder- and investor-genAI ensembles co-produce, contest, and recalibrate resonance in the rhetorical legitimation of new ventures. Building on scholarship in framing, micro-level legitimacy judgments, and sociomaterial affordances, I identify genAI rhetorical affordances (generativeness, extreme combinatorics, tone repertoire, velocity/energy and shared substratum) and theorize a recursive/iterative process model (ghost pitching, ghost screening, ghost relationship-building), configuring emergent resonance and legitimation. GFT builds new rhetorical framing theory for the age of genAI, connects research on human-AI collaboration with cultural entrepreneurship and extends affordance theory into multi-actor scenarios where affordance transitivity and visibility emerge as key considerations.

## INTRODUCTION

Gary Johnson III – co-founder of Delaware-based BVP Coffee Co. – was scrambling to enter the "Spirited Innovation Lab," a pitch competition whose USD35,000 grand prize could jump-start his whiskey-barrel-aged cold-brew venture. Facing "a plethora of questions to be addressed on the application" and a looming deadline, he turned to his "artificially intelligent assistant – ChatGPT" to co-create the entire application and then fit the "elaborate business plan into an illustrative 10-slide pitch deck" with it (Technical.ly, 2024). Was this a good idea? Would rhetorically framing entrepreneurial pitches conjointly with Large Language Models be conducive to resonance with key stakeholders, such as investors?

This real-world episode epitomizes a broader shift. A growing body of evidence confirms that "ghost" deployment of generative AI is no longer anecdotal. An investment fund study that blind-tested GPT-4 and human-written pitch decks with 250 US investors and 250 business-owners reported that 80 percent of respondents found the AI decks more convincing and were three times likelier to invest after reading them (Clarify_Capital, 2025). Another survey of 140 private-equity and venture capital firms revealed that 82 percent of PE/VC organizations were actively running generative AI modules inside their deal-sourcing and screening operations – up from 47 percent a year earlier. Interviews quoted in the same report describe a "shadow-AI" pattern in which junior associates quietly pipe pitch decks into language-model plug-ins for "Total Available Market checks", "competitive-grid autogeneration" and even "freshness scoring" narrative tone before a partner ever sees the file (V7, 2025). Taken together, these founder-side and investor-side statistics point to a rhetorical battlefield in which both sides increasingly suspect that the persuasive voice across the table may not be entire human.

Entrepreneurial scholars have long emphasized framing – the "use of rhetorical devices in communication to mobilize support and minimize resistance" (Cornelissen & Werner, 2014, p.

185) – as a primary route to acquire resources. As Lounsbury and Glynn (2001) first argued, new ventures must culturally position themselves through stories, analogies and prototypes that resonate with audiences whose attention is scarce. The framing literature has since blossomed into studies of category work, moral positioning, temporal narratives and the dark side of deceptive hype (Garud et al., 2025; Snihur et al., 2022). Yet, virtually all prior work presumes that framing is created and delivered by people, perhaps aided by basic presentation tools, but not by semi-autonomous systems capable of drafting entire pitch decks, generating photorealistic product mock-ups or simulating investor questions in a matter of minutes.

Generative AI (hereafter genAI) rewrites these assumptions in three ways. First, it affords near-zero-cost fluency: founders with modest writing ability can now produce grammatically perfect, stylistically polished decks that meet or exceed professional standards. Cognitive-ease research shows that fluent language often passes as more truthful and credible, even when substance is thin (Alter & Oppenheimer, 2009). Second, genAI enables the rapid recombination of expected pitch content elements (e.g., value proposition differentiation, market scenarios and revenue mode forecasting; otherwise arduous to produce), compressing a week-long brainstorming retreat into a ten-minute prompt-engineering session. Third, genAI authorship can be concealed, either deliberately or by omission – a useful feature, given the use of genAI appears to be heavily stigmatized (Reif et al., 2025).

These empirical developments motivate a fresh theoretical lens, which I will call Ghost Framing Theory (GFT). At its core, GFT contends that once AI systems co-author entrepreneurial narratives, the locus of agency in framing shifts from the individual entrepreneur to a human-AI ensemble, with ensemble defined as “using the complementary strengths of human intelligence and AI, so that they can perform better than each of the two could separately” (Choudhary et al., 2025, p. 543; Dellermann et al., 2019). What is more, evaluators, the target audience for entrepreneurial framing *also use genAI* for selection, feedback and ongoing relationship management (forming human-genAI ensembles *of their own*). The founder ensemble's rhetorical output intersects with evaluator ensemble processes in ways not well captured by existing theories.

I argue that genAI introduces five distinct rhetorical affordances – generativeness, extreme combinatorics, tone repertoire, velocity/energy and shared substratum – enabling both founder-AI and investor-AI ensembles to enact novel sets of rhetorical strategies in different stages (ghost framing, ghost screening, ghost relationship-building). These new-found strategies, and their interaction, in turn, require the theoretical restructuring of known evaluative process models of entrepreneurial framing and legitimacy (Bitektine, 2011; Snihur et al., 2022; Tost, 2011).

GFT makes several contributions. First, it re-centers framing scholarship on evaluator heuristics under AI conditions. I show that polish delivered by AI can increase resonance because founders draw on useful rhetorical affordances, inherent in genAI models. Second, I link hybrid human-AI agency research (Raisch & Fomina, 2025a) to institutional processes of legitimation, showing how ghost agency influences the legitimacy outcomes of venture communications. Third, I extend technology affordance theory (Leonardi, 2023; Leonardi, 2011) by characterizing genAI-specific rhetorical affordances and their transitivity and visibility, unfolding over time.

## ENGAGING THREE DIALOGUES

### Framing as Rhetorical Legitimation

Entrepreneurial framing research begins with the premise that new ventures lack the material track record and category-level scripts that established firms enjoy; they must therefore construct meaning and credibility in real time. Commonly with cultural entrepreneurship, narrative and rhetorical legitimacy (Lounsbury & Glynn, 2001; Zott & Huy, 2007), framing uses

"rhetorical devices in communication to mobilize support and minimize resistance" (Snihur et al., 2022, p. 578). As a point of specificity, framing 1) orients audiences' attention strategically, focusing on certain aspects with greater emphasis: "using language to strategically draw attention away from certain elements in order to draw attention to others" (Kalvapalle et al., 2024, p. 16) and 2) aims at resonance ("alignment with audiences' beliefs, values, and aspirations" as a proximate audience response (Snihur et al., 2022, p. 583), which is temporally more immediate and unit of analysis-wise more-micro level than new venture legitimacy as a subsequent and overall outcome (Fisher, 2020).

Several canonical framing tactics recur across this literature. Analogical framing links the venture to familiar exemplars (Cornelissen & Werner, 2014), such as the well-known "Uber for X" hook in entrepreneurial pitches. Emotional framing foregrounds affective cues such as founder passion or issue urgency to motivate stakeholders (Snihur et al., 2022). Categorical framing positions a startup within, or deliberately outside, existing industry categories to exploit legitimacy spill-overs and at the same time highlight distinctiveness (Ries, 2011; Navis & Glynn, 2011). Finally, temporal framing uses future-oriented narratives – road-maps, milestones, imagined end-states – to transform uncertainty into a sequence of achievable steps (Garud et al., 2014; Suddaby et al., 2023).

Yet, these insights rest on two tacit assumptions. First, that some messages are better framed than others – whether through structure, style or skillful refocusing – and this heterogeneity in framing quality makes some pitches more effective than others (Snihur et al., 2022). Audiences may assess the author's skill and draw inferences; well-framed messages thus reflect on the source (Nyilasy et al., 2025). Second, that authorship is transparent: the message is assumed to originate from known, accountable human agents, which can further contribute to positive resonance – a logic supported by empirical evidence in persuasion research, highlighting the effects of source credibility (Eagly & Chaiken, 1993).

GenAI unsettles both assumptions. While fluent narratives and crisp visuals can surely enhance resonance, when every founder can rely on these same rhetorical affordances, and with minimal effort, their diagnosticity for quality may diminish (Wingate et al., 2025). At the same time, while authorship grows opaque, reliance on intersubjective, collective wisdom (rooted in genAI's broad-based training) increases. The cost of idiosyncratic framing is the introduction of possible error and missteps – in clear contrasts with tried and tested patterns genAI can bring. These cracks in the fundamental assumptions of framing and rhetorical legitimation motivate the need for new theory: the concepts of ghost agency and rhetorical affordances, which reposition persuasive power within hybrid human-AI ensembles.

**Micro-level Legitimacy Judgements**

Legitimacy was long treated as a macro attribute – an invisible resource "owned" by organizations that behave appropriately within institutional logics (Suchman, 1995; Suddaby et al., 2017). Bitektine (2011) and Tost (2011) reconceptualized legitimacy as social judgments rendered by evaluators who operate under time pressure and bounded rationality. Central to this perspective is the distinction between propriety (individual assessments that the venture's actions are desirable) and validity (perceptions that there is a collective consensus of legitimacy about the venture). Individual judgments aggregate into collective legitimacy through social influence, media coverage and regulatory endorsement (Bitektine & Haack, 2015).

Tost (2011) deepened the legitimacy-as-perception view by identifying three content *dimensions* that underpin legitimacy judgements: (1) instrumental evaluations ("Will this venture create value?"); (2) relational evaluations (judging warmth and trustworthiness); and (3) moral evaluations ("Is it ethically right or fair?"). Judgments form through two *modes*: a passive heuristic mode dominated by cognitive fluency shortcuts (cognitive legitimacy) and a deliberative evaluative mode when the former is non-existent or comes into question.

GenAI places stress on assumptions about all three dimensions and both modes. By turbo-charging fluency, AI-authored decks can exploit the ease-equals-truth heuristic in the case of cognitive (taken-for-granted) legitimacy – and also inflate instrumental, relational and moral judgments in evaluative mode. Ghost authorship breaks this mapping, making it unclear whether smooth language reflects founder expertise, algorithmic assistance or hallucination. The proposed theory allows for a surprising and thus untheorized development: that *legitimacy evaluations themselves* may be imparted by genAI-human ensembles, rather than solo humans, redefining what socio-cognitive evaluation means in this new era.

**Technological Affordances**

The concept of affordances originated in ecological psychology, where they were defined as action possibilities the environment offers to an organism (Gibson, 1968). Organizational scholars imported the idea to explain why the same technology produces divergent outcomes across contexts. Early debates oscillated between material determinism and social constructivism; recent work argues for a relational view: affordances emerge from recurrent interactions between human intentions and artefact features (Bailey et al., 2022; Leonardi, 2023).

Even this relational scholarship carries an implicit assumption that an artefact's contribution is observable to users who "realize" the affordance. GenAI demonstrates that material contributions can be deliberately hidden, turning visibility into a strategic variable, not a contextual factor. For example, even if an entrepreneur is perfectly aware of the affordances ChatGPT offers to reshape, package and smooth over an entrepreneurial pitch, this affordance may vanish from view when hidden from evaluators providing funding for fear of perceptions of "cheating."

The proposed theory clarifies how interwoven genAI, founder and evaluator are in the relational process of ghost framing. GFT situates five rhetorical affordances – generativeness, extreme combinatorics, tone repertoire, velocity/energy and shared substratum – *on both sides* of the founder-evaluator dyad. By re-embedding affordances inside cognitive heuristics, it aims at synthesizing sociomaterial and evaluative perspectives. Doing so, it allows explanations not only on what genAI may enable but also how those new possibilities reverberate through the multi-dimensional legitimation process.

## GHOST FRAMING THEORY

To answer the challenges genAI poses for the above presented three dialogues (rhetorical framing, legitimacy-as-evaluation and affordance theory), I develop a new processual theoretical framework, I call Ghost Framing Theory (GFT) and build three theoretical propositions. In building the theory, I first introduce the notion of ghost agency as a conceptual metaphor borrowed from the practice of political ghostwriting. Second, I conceptualize five genAI affordances, which allow founders' and evaluators' framing strategies. Third, I outline the process model at the heart of GFT and propositions: the process of iterative and relational framing strategies by both founder and evaluator ensembles, during which framing resonance may (or may not) emerge.

### Ghost Agency

Most digital tools leave visible fingerprints – slide-deck templates betray Canva, spreadsheet macros hint at Excel – but state-of-the-art genAI can produce fluid prose[1] indistinguishable from human work (Peter et al., 2025). I define ghost agency as the general capacity of generative system-human ensembles (a) to create rhetorically meaningful utterances that are attributed, implicitly or explicitly, to a human actor, (b) drawing on large language models' training data and humans' experiences which form a shared substratum, (c) remaining partially or wholly invisible at the point of evaluation and iterative interaction.

[1] For simplicity, I will refer to textual output, though as the technology progresses, multimodal output (audio, imagery, video) is also becoming impossible to tell apart from human-generated content.

I use this conceptual metaphor of the "ghost" to refer to the practice of ghost-writing in politics, when an unseen, uncredited and often unknown writer helps a politician to deliver persuasive and rhetorically well-formed verbal utterances (Jamieson, 1990; Mapes, 2023). I consider this a pertinent metaphor, given that the use of genAI (similar to ghostwriters) is often effective, relies on a broader set of experiences than that of the speaker and yet it is anonymized and hidden for fear of stigmatization. I also use this metaphor to directly refer to the "ghost in the machine," a moniker used for AI more broadly (Leavitt et al., 2021).

Ghost agency, therefore, extends the notion of human-AI ensembles (Choudhary et al., 2025; Raisch & Fomina, 2025a) by foregrounding rhetorical and perceptual dynamics, which thus far received less attention, despite the widespread use (and stigmatization) of genAI in this context (Ling & Imas, 2025). Importantly, Ghost Agency complicates prevailing assumptions in entrepreneurial framing and new venture legitimacy. If evaluators (such as investors) subconsciously or consciously equate communicative polish with domain competence, then hidden genAI contributions constitute a form of rhetorical leverage unavailable to earlier cohorts of entrepreneurs. Within this unsettled context, one must specify exactly what genAI affords actors to do differently in rhetorical legitimation processes.

**GenAI Rhetorical Affordances**

Building on affordance theory (Gibson, 1968; Wang & Tracey, 2024) and emerging literature on genAI affordances (Grodal et al., 2024; Phillips et al., 2024), I identify five affordances that genAI offers to entrepreneurs and evaluators for framing and rhetorical legitimation. In agreement with the relational, interactive and iterative understanding of entrepreneurial framing (Garud et al., 2025; Snihur et al., 2022), I will conceptualize each of the five affordances sharing commonalities between the two sides of the founder-evaluator dyad. For ease of presentation, I first enumerate these affordances and in later sections I theorize how entrepreneurs and evaluators may rely on them in their rhetorical framing strategies. This separation, however, is only a rhetorical device: I acknowledge that they are deeply intertwined in human-genAI ensembles (Choudhary et al., 2025; Leonardi, 2023). Table 1 defines the affordances and details genAI features supporting each of the affordances.

----- Insert Table 1 about here -----

***Generativeness.*** The most immediately striking affordance of genAI is its ability to produce large amounts of rhetorical content, seemingly “out of nothing.” Indeed, the genAI label itself refers to this general affordance, which is also often used to differentiate it from other forms of artificial intelligence, such as “predictive AI” (Davenport & High, 2024; Hermann & Puntoni, 2024). This basic affordance is likely the most alluring of all for human users who often need to produce sizeable amounts of linguistic material to achieve goals. Importantly, and in contrast with predictive AI, generativeness does not guarantee reliability and validity, as the often highlighted “hallucination” phenomenon (untrue or irrelevant content) testifies (Gosmar & Dahl, 2025). This is distinctive from predictive AI, where reliability and validity are core expectations.

***Extreme combinatorics.*** GenAI is not only able to generate things out of nothing. It is also capable of taking inputs (such as strategic documents, hastily typed rough ideas, personal emails, visuals and process charts, formal technical requirements and task specifications, etc.) and turn them into fluid prose. While the combinatorical nature of creativity (“new combination of old elements”) has long been observed (Harvey, 2014; Mayer, 1999; Osborn, 1961), humans have very real cognitive limitations in arranging “old elements” into “new combinations” – especially, when there are a lot of them and when it is highly uncertain what combinations would be appropriate. In contrast, genAI has vast resources to rely on in this domain, only seemingly constrained by the number of its training parameters and the size of the query’s allowable context window – but even small models are startlingly capable (Kumar, 2024).

***Tone repertoire.*** GenAI is versatile in stylistic expression giving the affordance to users to vary tone in seemingly infinite amounts of registers, as circumstances require. This affordance is supported by a surprising observation about genAI: that it does not behave like mechanistic computer software (Mollick, 2024). Unlike other forms of related information technology, genAI is a remarkably skilled and humanlike writer. Indeed, the anthropomorphic ability to take on different personas and adjust stylistic tone considering various audiences was one of the features that both mesmerized and shocked early users of the technology (Zao-Sanders, 2025). In entrepreneurship, this affordance is both a great opportunity to aid rhetorical framing and also a threat that may lead to deception (Eliot, 2025).

***Velocity/energy.*** Fourth, genAI is relentless and affords astonishing speed and energy in accomplishing tasks once believed exclusively human (and thus subject to human limits in cognitive processing and motivational energy). Compared to humans, genAI is capable of producing vast amounts of output in short amounts of time – and it is not subject to depletion. This affordance is critical in management contexts where expectations are often high and available resources limited. Thus, not surprisingly, knowledge workers in various industries were fast to adopt the technology to boost their individual productivity – even when organizations disallowed such use (Felten et al., 2024). The affordance of limitless energy and velocity is, in a general sense, is a great allure to resource-scarce entrepreneurs.

***Shared substratum.*** While less visible, genAI also affords access to a vast pool of shared human experiences. This affordance is rooted in the very nature of genAI's training, the fact that it relies on large corpora of human generated text (Vaswani et al., 2023). Indeed, the scale of this training is remarkable. The original training of ChatGPT-3, one of the first, now outdated models, contained "Common Crawl" (a webscrape of large portions of the Internet), the entirety of Wikipedia, two large corpora of scanned books and another Internet source – altogether nearly 500 billions of tokens (a common measure of training size; Brown et al., 2020). Thus, genAI affords unprecedented access to the shared experiences of humanity (even if represented in an unreliable and biased manner; Doshi et al., 2025), which no individual can hope to achieve based on their own human "training" (formal education, media use, on-the-job training, everyday experiences and so on).

**The Ghost Framing Process**

Figure 1 visualizes GFT as a relational and recursive cycle that unfolds over time between two human-genAI ensembles (founder-genAI and evaluator-genAI). The process starts with initial transactional exchanges (ghost pitching and ghost screening). If resonance (or perceived fit between venture narrative and evaluator legitimacy judgments, Benford & Snow, 2000; Tost, 2011) allows, the process morphs into iterative relationship development cycles (ghost relationship-development).

----- Insert Figure 1 about here -----

When deploying rhetorical strategies, entrepreneurs now have the affordances of genAI systems available to complement their own skills. Naturally, entrepreneurs would engage in rhetorical framing even without genAI – and may very well succeed without the use of the technology, as they had before its appearance recently. As my first proposition, I will argue, however, that the use of GFT can differentially improve new ventures' chances for framing resonance because genAI affordances[2] enable novel and effective framing strategies (holding everything else, such as underlying venture quality, constant) and diminish idiosyncratic error.

My second proposition is that this outcome is contingent on whether the evaluator ensembles themselves also use genAI for evaluation. If founders did not use genAI, ghost screening will increase the likelihood of being screened out; conversely, the positive effects of founders' use of genAI for framing are strengthened by evaluator-genAI ensembles.

My third proposition is that, tracing the process further over time, if the venture survives initial screening and segues into the relationship-development stage, the interactive effects of founder and evaluator genAI use become less important, the above contingency is ameliorated by the developing human relationship between founder and evaluator, turning genAI use by either partly more uniformly positive. Figure 2 summarizes the three core theoretical propositions expanded upon in the subsequent three sections. Table 2 summarizes founders' and evaluators' ghost strategies.

----- Insert Figure 2 about here -----

----- Insert Table 2 about here -----

***Ghost pitching.*** In this initial pitching stage (Kalvapalle et al., 2024), founder-genAI ensembles aim to establish a clear and persuasive narrative to secure initial support ("a foot in the door") from key stakeholders in the process of establishing medium- or longer-term resonance. As the sections below will establish, these genAI-affordance-enabled pitching

[2] Below, I will align these affordances with the most likely strategies they enable, pairing the two most important affordances with each strategy. I acknowledge that some strategies are aided by more than two. However, in line with theorization in the past on affordances, I find it useful to highlight which affordances help which strategies the most, even if other combinations are also possible (Wang and Tracey, 2024).

strategies are expected to positively influence resonance/legitimacy outcomes.

> ***Proposition 1 (Ghost framing effect)***: Controlling for underlying venture value (costly signals), founders using ghost framing strategies (enabled by genAI affordances) are more likely to resonate with evaluators in the initial stages of legitimation, than those who don't.

At this early stage, founders themselves are rather unclear about the core parameters of competition, the nature of the opportunity or their own capabilities. Therefore, framing strategies serve as much as gaining favors from make-or-break audiences as persuading themselves – a process that often takes a strong emotional toll (Neuberger et al., in press). GenAI affordances promise to provide some relief amidst this highly uncertain and stress-prone context. Those entrepreneurs relying on genAI deploy three novel framing strategies: (1) robust scaffolding, (2) multidimensional balancing and (3) context expansion. The strategies are novel in that they intertwine human skill and non-human technological affordances in ways not seen before genAI.

*Robust scaffolding* refers to entrepreneurs using genAI when creating rhetorical structures in evaluator-facing communications that are both plausible and easy-to-digest. As entrepreneurs face extreme uncertainty, they constantly face the risk of (1) either misconstruing key features of the opportunity or their own capabilities or (2) miscalculating the best approaches to convey their competitive advantage to evaluators. The strategy involves deploying imperfect but temporarily helpful rhetorical structures to "scaffold" evaluators, guiding the comprehension of the business model and the demonstrating its value.

While non-gen-AI methods are also available to achieve this (such as general resourcefulness, advice, mentoring or paper-and-pencil cognitive aids such business model "canvassing"; Osterwalder et al., 2010), genAI represents a step-change. The affordance of *generativeness* means that, simply, genAI can seemingly do it all: It can create a convincing-sounding business model, it can create financial forecasts and it can pull together a plausible pitch deck, seemingly out of nothing. The *velocity/energy* affordance enables entrepreneurs to "jump hoops" and "get out of the doldrums" when, otherwise, founders may feel desperate and when there is no help available. The relentless energy of genAI to generate content (no matter what the problem is) can momentarily "unstick" the framing process – which may be helpful, even if the solutions in the end are hallucinations.

For example, an entrepreneur may be able to automatically create a frame for a financial model when no clear path to profitability exists or phrase value propositions more realistically and concretely. Founders may also be able to scaffold audience understanding by structuring a pitch deck with "winning" new venture templates such as those related to Business Model Canvas, Design Thinking or Disciplined Entrepreneurship. While these templates are available for use without genAI, their application in the particular context of the venture may be challenging still. GenAI-based tools, such as those of GPTHacks.com offering detailed advice and prompts how to build pitch decks, however, make such audience scaffolding tasks easy.

*Multidimensional balancing* is a strategy that involves the effective resolution of the possible contradictions between either the different elements of the new venture's business model or amongst its set of assumptions. While such balancing has always been part of "finding a narrative," which gives coherence and glosses over possible conflicting elements in the venture concept (e.g., between social impact motive and financial reality), the novelty is the ease, speed and depth of accomplishing this balancing across a high number of factors and criteria (hence the label "multidimensional"). Indeed, it is commonsense practice now to upload large amounts of text (e.g., consumer insights, financial models, competitive information, supplier needs, etc.) to ChatGPT to reconcile them into the semblance of a plausible business model.

The affordance of *extreme combinatorics* allows the simultaneous consideration of seemingly incommensurate pieces of information or interests of diverse stakeholder groups. More broadly, it involves the reconstitution, the creative assemblage of prior fragmented ideas into coherent rhetorical wholes. It seems that these affordances dramatically increase the dimensionality of factors and criteria considered: from a few variables a human can balance to possibly tens (or even hundreds) genAI can. The affordance of *velocity/energy* makes this high-dimensional optimization readily available – compared to relying on human cognition and motivation, which are subject to strict limits and fast depletion.

An example of multidimensional balancing is the quick resolution of conflicts of interests between two legs of a platform business model. Advertisers and end-consumers for a lifestyle app (e.g., giving healthy eating advice) may have a long list of only partly overlapping preferences. GenAI may be able to parse preferences rapidly and frame mutually beneficial preferences as a story of a “win-win,” supported with actionable recommendations. Another example may be the selection and marketing of optimal features for a Minimum Viable Product for the favored “beach-head target” – selected based on the inputs of highly technical feature documentation and seemingly incommensurate qualitative consumer interview transcripts.

The third strategy is *context expansion*, discovering and integrating into the pitch diverse perspectives, ways of thinking or modes of self-expression that are different from that of the founders – but useful in gaining resonance from evaluators. Commonly, founders are highly focused on their startups and the day-to-day activities associated with them – at the extreme, to the point of obsessive passion (Cardon et al., 2009). A side-effect of this intense focus is the possible disregard of other stakeholders’ perspectives. A common observation of founders’ product-orientation (focusing on technical features) rather than consumer-orientation (understanding consumer preferences) was one of the key motivators of the emergence of lean startup (Ries, 2011), a movement aiming partly to expand founders' narrow focus. Another common problem is favoring the value proposition (what the venture offers) too much at the expense of financial modelling (how the venture makes money) (Aulet, 2013).

In genAI-assisted pitching, founders can readily overcome these limitations. Founders can adapt ideas across diverse stakeholders and settings, offering a wider lens than any one individual may have. It can involve the imputation of elements of knowledge, references, or stylistic flair that exceed the entrepreneur's initial input. GenAI can easily be instructed to take on different "personas," prototypical human occupational or other group memberships (Mollick, 2024). This ability is rooted in the fact that LLM models are trained on diverse sets of textual material, with linguistic utterances, perspectives and ways of thinking coming from "all walks of life." Therefore, when prompted, these implicit perspectives can be evoked.

The most relevant affordance is *tone repertoire*, that it can represent varying social groups' linguistic style with surprising ease. This stylistic matching between founders and evaluators can clearly aid resonance (Teeny et al., 2021). Also, it is *shared substratum,* the vast repository of intersubjective human experiences that makes it possible for founders to discover symbolic regimes that may even be unknown or unpalatable for them otherwise.

Several examples can be evoked for the context expansion strategy. Founders uncomfortable with finance can now create a financial model section in a pitch deck that seemingly speaks the "language of finance" flawlessly for VC audiences. This can increase instrumental legitimacy amongst these audiences signaling investability. Similar stylistic matching can be useful in highly bureaucratic environments, such as applications for government grants, where expectations for linguistic precision can be exacting (and unfamiliar for founders).

Also, communicators can adopt the language and thinking patterns of little understood consumer target groups and, in turn, showcase their "empathizing with the consumer," a valorized practice in relevant evaluator communities of practice (e.g., in "design thinking," Micheli et al., 2019).

In sum, I theorize ghost pitching to have a positive effect on resonance because it enables founder-genAI ensembles to transform nebulous venture concepts into investor-ready artefacts by swiftly scaffolding coherent narratives, balancing conflicting business-model elements and importing audience-specific perspectives. These genAI-driven capabilities let entrepreneurs iterate, tailor and clarify their pitches with unprecedented speed, winning early stakeholder attention while sharpening their own strategic understanding.

***Ghost screening.*** Today many evaluators at this initial stage engage in what I label ghost screening, the use of genAI affordances to aid with venture selection decisions (Zhang et al., 2024). It has long been recognized that evaluators of entrepreneurial rhetorical legitimation attempts are not mere recipients – instead active processors and framers of the utterances (Huang, 2018). This picture now is further complicated by the fact that many investors also rely on genAI assistance for evaluation, which appears to help them (Doshi et al., 2025). Therefore, evaluators may form human-AI ensembles, taking advantage of the same genAI affordances as entrepreneurs do – but for different tasks. New theorizing is needed because the involvement of non-human agents breaks assumptions of the evaluator sense-making process – namely, that is rooted in exclusively human expert judgment.

My core prediction is that the use of genAI affordances facilitate a new set of framing strategies among evaluators, which in turn positively moderate the formation of framing resonance (or dissonance) between founders and them.

> ***Proposition 2 (Ghost screening moderation)*****:** The ghost framing effect (P1) is moderated by ghost screening such that (a) the positive effect of founder genAI use is strengthened when evaluators use ghost screening, and (b) when founders do not use genAI, evaluator screening has a negative effect on resonance.

These screening strategies differ from those of entrepreneurs because evaluators take different positions in the legitimation process. Naturally, their roles are the processing, reframing and ultimately legitimating of frames created by the entrepreneurs – not the origination of these messages.

However, there are important commonalities as well. Evaluators are not merely passive "consumers" of entrepreneurial framing (Nyilasy et al., 2025). They actively make sense, reframe and recommunicate these messages to various audiences (internal and external) – as founders do, too. Importantly, evaluators shape the overall frame resonance and rhetoric legitimation process itself, by playing a key role in its social construction (Berger & Luckmann, 1966). Another common element with entrepreneurs is the fact that they may rely on the *same genAI technologies* as used by entrepreneurs. This has important consequences. Resonance in this new technological setting is not merely influenced by entrepreneurial rhetorical skill or evaluator's human cognitive processing and sense-making. GenAI, if used by both parties (as likely to be the case today) establishes a common technological ground between them. This common ground, as manifested by shared technological affordances in the ensuing ensembles, may be a direct (and novel) contributor to emerging resonance. Next, I will theorize three core ghost screening strategies that evaluator-genAI ensembles enact when processing and shaping entrepreneurial rhetorics: rapid vetting, speed distillation and power feedback.

*Rapid vetting* is a ghost screening strategy where evaluators, such as investors process large amounts of rhetorical pitch documents relying on the affordances of genAI. The motivation for involving genAI may purely lie in the volume of work submitted, which is often impossible to reliably process without machine help. But the intention may go beyond quantity and involve quality: an artificial intelligence system may be perceived as more standardized and reliable, especially, if it is based on Retrieval-Augmented Generation (RAG), i.e., complementing off-the-shelf genAI systems with specific criteria captured in documents supplied by evaluator users. Importantly, the reliance of genAI does not necessarily mean “outsourcing” the entirety of the evaluation process, which evaluators may find unpalatable for various reasons (perceived accuracy or social identity maintenance). It does mean the incorporation of genAI into evaluator-genAI ensembles for at least a proportion of the process.

The key affordances this strategy relies on are *extreme combinatorics* and *velocity/energy.* Extreme combinatorics allows highlighting multidimensional criteria to be balanced, with the expectation of making a meaningful prediction: in this case, if a pitch proposal should be screened out or screened in, for further investigation. *Velocity/energy* is an affordance relied on greatly with this strategy; indeed, were it not for the large volume of pitches, evaluators may wish to retain autonomy over evaluation entirely. GenAI is tireless and consistent, and if clear criteria are communicated it is capable of mitigating human limitations, including cognitive depletion or simply, labor availability.

An example is Antler, a global VC, which relies on Persona Studios’ genAI assistant to screen large amounts of founder applications each year. Antler has tested thousands of pitch decks using this system and even came up with heuristics based on these tests, which they label “Spike Theory”: looking for top-decile excellence in any one of the selection criteria (Persona_Studios, 2024). GenAI thus enables evaluators not simply to make decisions, but also to make sense and define rhetorical new venture legitimacy.

The second strategy, *speed distillation* is the summarizing of entrepreneurial narratives with the purpose of formulating actionable recommendations for screened-in ventures for internal evaluator audiences. While the screening process has a direct effect on entrepreneurs, it also has a life of its own inside screening organizations. First, rarely does the process of screening mean a singular "invest vs. don't invest" or "accept vs. don't accept" decision. Instead, screening is simply the first step in a multistage process, which involves much ongoing deliberation (Zhang et al., 2024). Second, screening itself has multiple audiences (in the case of this strategy: different internal groups within the evaluator organization), which are involved in evaluation decisions. Speed distillation relies on genAI to partially automate this process for greater consistency and alignment with the evaluator's internal preferences. Importantly, this strategy highlights that evaluators do not simply consume entrepreneurial framing, they actively shape it and reproduce it, using their own framing – which then contributes to the overall emerging rhetorical legitimation process.

The affordance of *generativeness* supports this strategy through the autonomous production of internal narratives, for example, by writing memos or reports summarizing strategic recommendations for a possible investment target. The fact that genAI is capable of coming up with such documents with minimal input mirrors the generativity of human-entrepreneur ensembles. *Shared substratum* is helping this strategy given the multifaceted nature of evaluator sense-making in this domain. Criteria are manifold in uncertain new venture settings and humans have well-defined limits in how much information they can reliably process. Relying on vast repositories of relevant intersubjective experiences contained in genAI's training, human-genAI ensembles have the promise to overcome these limitations (De Freitas et al., 2025).

An example is Flybridge Capital Partners, who built an AI-powered “Investment Memo Generator” that ingests a pitch deck or transcript, runs market-research “agents,” then auto-drafts the standard “Why invest?” memo, including market size, team, business model and risk evaluations (Reyes, 2025).

*Power feedback* is a strategy that involves providing rhetorical responses to founders, at scale, with content and tone relevant for the communicated pitch proposal. While evaluators such as VCs may have always provided feedback, with the advent of genAI this activity has shifted: it became much larger in volume and likely changed in substantive content and style as well (possibly being more consistent, while also probably more isomorphic). To contrast with the previous strategy, speed distillation, power feedback has both a different (1) target audience (entrepreneurs rather than internal stakeholders) and (2) purpose (attempts to explicitly influence founders’ strategy).

Two affordances are critical for this ghost screening strategy: *tone repertoire* and *velocity/energy*. Tone repertoire means the ability to generate feedback matching the founders’ frames generated. Evaluators may find it important to provide resonant feedback to founders, especially, if they see potential for future collaboration. From that perspective, tone repertoire is a critical affordance since it is stylistic matching, a known driver of resonance (Ireland et al., 2011). Second, the affordance of velocity/energy replaces the impossibly large amount of time it would take to provide feedback using manual labor. This possibly increases both the scale and depth of evaluators’ search activities.

For example, Ada Ventures’ AdaGPT provides instant feedback on founder pitch decks in structured areas: investment thesis alignment with the fund, venture potential (innovation, competition, market opportunity, scaling), team, developmental areas and next steps

(Ada_Ventures, 2025).

In summary, in ghost screening, evaluator-genAI ensembles actively process founder pitches rather than merely “receiving” them, using LLM affordances to triage large volumes of material, surface inconsistencies, and recast the narrative for their own decision making. Through rapid vetting, speed distillation and power feedback, investors automate memo writing and tailored responses, creating a shared technological ground on which both sides co-construct frame resonance – or expose dissonance – at unprecedented scale and speed. Ghost screening modulates the effect of ghost framing: it strengthens the positive effect of founders’ genAI use, but also makes non-users more likely to be screened out. As I argued, this moderative effect is explained by the shared use of affordances, most importantly the shared substratum, on which both founder-genAI and evaluator-genAI ensembles may rely.

***Ghost relationship-building.*** The process of new venture legitimation does not stop with pitching (Kalvapalle et al., 2024). A relational perspective (Bailey et al., 2022; Huang & Knight, 2017) has been adapted to fully capture the interactive and co-constitutive nature of evolving rhetorical legitimation over time (Neuberger et al., in press; Snihur et al., 2022). To signal the extreme high velocity and conjunctive nature of genAI-assisted relationship development, I refer back to the visual metaphor of the dampened harmonic motion of a coiled spring (Figure 1), indicating many loops of fast-paced interaction, moving in the direction of increased resonance.

With the advent of genAI affordances, founders and evaluators can now build these relationships faster and more efficiently than in the past – but more importantly the interactive effects of ghost pitching and screening dissipate over time, as the relationship develops. Formally, this can be stated as a three-way interaction:

> ***Proposition 3 (Relational alignment effect)*:** Ghost pitching strategies, ghost screening strategies and processual (initial vs. relationship) stage interact in a three-way interaction on resonance/legitimacy outcomes, such that ghost screening moderation effect (P2) dissipates along the process of relationship development.

I will first present the effects of founder framing strategies (hyperpersonalization and enhanced responsiveness), followed by evaluator strategies (proactive surveillance and bias mitigation), acknowledging the order is arbitrary, given the iterativeness.

The first founder strategy is *hyperpersonalization.* While even first shot attempts at ghost pitching may contain elements of personalization which genAI can aid at scale, it is in the ongoing relationship management stage (when founders have gotten initial buy-in from evaluators and they continue to engage with them over time) where true personalization emerges. This is because the ongoing and iterative nature of the relationship allows deeper understanding of each party's needs. While this was certainly the case even before (Huang & Knight, 2017), the difference in the age of genAI is (1) that the core difficulty with customization, resource intensiveness (in time and energy) of personalizing persuasive messages can easily be overcome and (2) that this can be done at a greater level of granularity, down to the level of word selection (Berger et al., 2020).

The key affordances that help make this happen are tone repertoire and velocity/energy. *Tone repertoire*, the affordance rooted in the fact that genAI is capable of producing a vast array of possible voices in every imaginable linguistic context, allows even unskilled communicators to readily adopt a persona with a communicative style that matches evaluators' needs. *Velocity/energy* is the affordance, in turn, needed to implement customization against high resource needs. Since startups generally are resource poor (in time, energy, money), the use of genAI can effectively resolve the trade-off between greater effectiveness through customization vs. greater resource needs.

Examples include uploading profiles or even full audio or text transcripts from known audience members to “mirror” their style or preferences. Indeed, genAI-based software-as-service tools often position themselves on this basis. For instance, Visible outlines how genAI tools can help customize investor updates, generating “updates tailored to specific investor groups or individual stakeholders based on their interests or investment focuses” (Graumann, 2023)

*Enhanced responsiveness* is a founder strategy to increase the perceived speed of addressing evaluator requests. Digital technologies have probably generally sped up response times in founder-investor communications, however, genAI technologies can make iterative communication loops almost instantaneous. The perception of responsiveness, in turn, results in greater alignment both because of increasing trust but also because possible misunderstandings between evaluators and founders can be ironed out quicker. Further, founders can effectively stage a “coachable” founder persona, a known evaluator preference (Nyilasy et al., 2025).

The affordance of *shared substratum* is of key value here: it enables founders to reduce uncertainty based on the coherence intersubjective training data may provide. With high uncertainty, it is extremely difficult for founders to react to varying (or even conflicting) requirements (Neuberger et al., in press). Evaluators themselves may not be fully clear about their own expectations (Zacharakis & Meyer, 1998). Thus, relying on an intersubjective set of shared expectations (contained in the training data) reduces uncertainty, and aids resonance. Second, *tone repertoire* makes it possible to stylistically match founder language used to that of evaluators. This affordance increases a semblance of receptiveness even when the underlying response content is less than perfect. Under such circumstances, a stylistically fluid response is of high value.

As an example, Inari, an AI customer relationship management startup in Y-Combinator's Summer 2023 intake built a private GPT that asks a few contextual questions (monthly recurring revenue [MRR], churn drivers, hiring needs) and then auto-generates a ready-to-send monthly update (Inari, 2025). This tool demonstrates the utility of genAI affordances sending signals of enhanced responsiveness, otherwise prohibitively arduous at high speeds by merely relying on human labor.

Turning to the evaluator side, *proactive surveillance* is a strategy of anticipating and preventing likely problems that may derail startups, with an eye for early-stage interventions. Post-investment monitoring is notoriously difficult for investors, given the small scale, private ownership and contextual uncertainty involved (Gompers et al., 2020). With genAI, investors can anticipate common startup problems, specific to venture stage development, industry context, and market information – and then proactively query founders on these points.

The affordances that make this possible at scale and at high speeds are extreme combinatorics and velocity/energy. *Extreme combinatorics* facilitates probing of inconsistencies and gaps in information provided by founders on an ongoing basis. While investors may believe they have good gut instinct for driving portfolio ventures to high performance (Huang & Pearce, 2015), with large volumes of information and uncertainty it is easy to miss indicators of trouble. *Velocity/energy* allows coping with the volume, inherent complexity and uncertainty of founder information – or that of the market and competitors, all evolving over time.

For example, SignalFire an AI-powered venture capital firm promises investors the ability to proactively surveil its portfolio of invested ventures, looking for early signs of trouble and devising preemptive strategies: SignalFire "doesn't just notify investors when a founder is about to leave a unicorn. It flags red flags in GTM motion, predicts breakout potential, and integrates directly with portfolio company workflows" (Mathews, 2025). Even when evaluators do not have specialized in-house tools, they can generate simple genAI prompts to similar effects.

Finally, *bias mitigation* is an evaluator strategy that controls known or unknown evaluative decision-making biases by relying on genAI. The strategy is based on the fact that genAI is exposed to a broader spectrum of experiences than any one evaluator. Therefore, its use can limit the subjectivity inherent in any one individual's decisions. This may indeed be valuable, given that new venture investors are often overconfident and investment experience has diminishing returns (Shepherd et al., 2003). While genAI itself may have biases rooted in its training (Atari et al., 2024), in the context of probing startups' about their progress in the portfolio of investments, the use of genAI could allow investors to achieve greater alignment with these startups if they are able to mitigate some of their decision-making biases.

In terms of affordances, *generativeness* allows frequent and automated communication, which is devoid of idiosyncratic biases. The mere fact that genAI is capable of coming up with generally plausible utterances for most scenarios means that, if used at scale, it would dilute any bias rooted in any one individual (though systematic biases are possible, as shown above). *Shared substratum* provides access to large-scale intersubjective experiences, which controls for common decision-making biases, an effect that can be further enhanced by targeted prompt engineering on this point (as argued later).

As an example, the Emerging VC Accelerator adapted genAI-based DebunkBot (originally devised for conspiracy theories; Costello et al., 2024) to control for biases native in venture capital. For example, confirmation bias is so engrained that even seemingly objective due diligence processes fall prey to it: "When VCs allow early impressions to dominate, they risk falling into the trap of confirmation bias, selectively gathering information that supports these initial beliefs rather than challenging them. In these cases, even rigorous DD [due diligence] can be skewed, as the data collection itself becomes a biased search for validation" (Attar, 2025). As DebunkBot has proven, genAI, due to its broad-based training, can be used to overcome biases such as this.

In sum, once initial buy-in is won, gen-AI enabled founders and investors may enter fast-spinning "coiled-spring" exchanges in which genAI powers hyperpersonalization, enhanced responses, proactive surveillance and bias mitigation. Consequently, the conditional effects of framing and screening I postulated in the initial stage transform into a singularly positive one, because genAI-boosted relationship strategies let each side anticipate the other's needs, close misunderstandings quickly and co-evolve a resonant narrative that sustains legitimacy over time.

## DISCUSSION

The emergence of genAI in the new venture framing and legitimation process created a new reality for both founders and evaluators. The consequences of both founder- and evaluator use can be far-reaching as genAI has been the fastest adopted technology in human history. It certainly was consequential for Gary Johnson III, from the opening vignette, who ended up winning the competition – even after openly declaring his reliance genAI affordances (Technical.ly, 2024).

Yet, this important technological transformation and related entrepreneur/evaluator processes have not been built into theories of new venture legitimation. I have developed Ghost Framing Theory to overcome these shortcomings by amalgamating relevant scholarship on technological affordances, new venture legitimacy and entrepreneurial framing. In my discussion section, I review three core contributions and avenues for future research, which includes the investigation of possible boundary conditions.

**Framing and Rhetorical Legitimation**

Entrepreneurial framing and rhetorical legitimacy scholarship has shown how founders may mobilize analogies, categories and narratives to secure resources under uncertainty (Snihur et al., 2022; Zott & Huy, 2007); yet, it has not had the chance yet to examine how those rhetorical formations exert their influence *once machines assist either (or both) sides of the exchange*. Even in the AI era, it has is often assumed that rhetorics is an inherently human activity, ill-matched with machines and artificial intelligence; AI is appropriate for mechanistic tasks, inappropriate for creativity and invention, which are uniquely human (Castelo et al., 2019).

The appearance of genAI, a unique type of artificial intelligence, seems to have defied these long-held expectations (Mollick, 2024). It appears genAI may, in fact, be better at humanlike creative endeavors such as creating rhetorical texts, than machinelike, consistent prediction (Hermann & Puntoni, 2024). Somewhat surprisingly, genAI behaves less like software, and more like a human when interacting with human prompters (an arrangement I labeled genAI-human ensembles). These ensembles, in turn, are able to conjure up humanlike creative content (and thus entrepreneurial framing) not only matching pure human skill, but also potentially surpassing it (Lee & Chung, 2024).

GFT also extends the evaluator-focused paradigm of legitimation (Bitektine, 2011; Tost, 2011) by explaining why this is. The core proposition (P1) of the theory is that genAI-enabled entrepreneurs are better at achieving resonance with evaluators than those without because genAI affords unique skills and useful strategies built on these affordances. A key affordance is accessing the *shared substratum* (the large intrersubjective pool of human experiences captured by genAI's human training data), which resonates with audiences given that a) it is also *their* shared human experience and b) because it reduces the idiosyncratic error each framing attempt may contain.

Going further, GFT builds a recursive, longitudinal model where *evaluator*-genAI ensembles modulate this core process (P2): enhancing the positive effects of founder genAI use and penalizing non-use. I also argue that over time this very conditionality diminishes and turns dominantly positive (P3). This is because continued genAI reliance (this time for relationship building strategies) results in an ever-increasing spiral of iterative, co-constitutive "resonance-making," a form of mutual sense making in the context of an evolving relationship.

**Human-AI Collaboration Research**

Studies of human-AI collaboration emphasize the crucial issues of automation-augmentation (Raisch & Krakowski, 2021) and cognitive complementarity in organizational problem solving (Raisch & Fomina, 2025b), while rhetorical legitimacy scholars examine how actors create or disrupt norms through symbolic work (Suddaby & Greenwood, 2005).

My model bridges these largely disparate scholarly domains by theorizing hybrid agency *within* symbolic legitimation. When founder-genAI ensembles co-create pitch decks, agency becomes layered: human strategic intent, LLM linguistic execution and even investor-side genAI evaluation intertwine. This layered agency reshapes classic legitimation dynamics. Work in new venture legitimacy predicts firms to mimic or link to peers, trusted institutions and categorical conventions to secure legitimacy (Fisher, 2020; Lounsbury & Glynn, 2001). GFT accelerates this mimicry by templating rhetorics deriving from a vast repository of intersubjective cultural material inherent in genAI's training data. Thus, my theory demonstrates that human-genAI ensembles are capable of symbolic and not merely functional work, which is a significant development that needed new theorizing.

In doing so, my model fundamentally assumes genAI engages in human augmentation rather than automation (Raisch & Krakowski, 2021; Raisch & Fomina, 2025a. There are two main reasons supporting this assumption. One is inherent in the technology itself. GenAI is technologically different from other types of AI in that it is fundamentally dyadic, dialogic and co-creative. Indeed, it would be impossible to achieve any meaningful output without the active participation of the human prompter whose information and skill radically shapes what "comes back" in a two-way conversation with genAI models. "Agentic" models, a new class of AI models just emerging (which would aim to prefer automation over augmentation) may rewrite this assumption (more on this in Future Research below). Second, entrepreneurial framing is hard to theorize to be anything but co-constitutive. Entrepreneurs are known to be highly passionate (Cardon et al., 2009) and agentic (McMullen & Shepherd, 2006). It would be hard to believe that they would leave crucial framing strategies to purely automated agents in such an important activity as rhetorical framing for seeking resources and legitimacy (Zott & Huy, 2007).

In short, GFT complements emerging work on human-AI collaboration in organizations by theorizing how genAI augments entrepreneurs in their efforts to gain resonance and legitimacy, also tracing how this process unfolds over time and by the co-creative participation of evaluator-genAI ensembles.

**Affordance Theory**

Affordance scholarship (Leonardi, 2023; Wang & Tracey, 2024) has illuminated how material artefacts enable or constrain action, but empirical work has skewed toward physical or information-processing tasks. I introduce genAI rhetorical affordances – action possibilities that shape entrepreneurial framing attempts.

I move affordance research forward in three ways: through their genAI rhetorical affordances' *specificity* (they are particular to their action possibility domain, rhetorical framing in entrepreneurship), their *transitivity* (the recognition that they ultimately enable a third party, evaluators, e.g., VCs, making a funding decision) and their *visibility* (they may or may not be known by evaluators, hence the Ghost Framing moniker).

GenAI rhetorical affordances (generativeness, extreme combinatorics, tone repertoire, velocity/energy, and shared substratum) are indeed special in that they capture how an emerging technology enables novel entrepreneurial strategies and action. I also demonstrate that these affordances are unique in that they are transitive: they enable not only the founder, but also the recipient evaluators, who ultimately benefit from engaging with a well-pitched venture idea. Finally, I highlight affordance visibility, which becomes especially interesting when a third party is involved since observing the genAI-founder enablement is not trivial. Further, given the stigmatization of genAI use (Reif et al., 2025), affordance visibility may be a feature founders wish to strategically control.

**Directions for Future Research**

GFT opens up several avenues for future research, both in entrepreneurship and organizational scholarship more broadly. First, testing important boundary conditions can extend the model. On the level of the individual, variation in prompting skill may make a crucial difference. Since genAI-human ensembles are, by definition, co-constitutive, what humans bring to the prompting "table" may have significant negative or positive consequences. For example, AI generated content may lead to homogenization (Hsu & Bechky, 2024), and form "AI slop" (Elgan, 2025). "One shot prompting," typing in simple instructions without much thought, context or iterative engagement, may make outputs less unique, undermining competitive advantage. Conversely, entrepreneurs skilled in "prompt engineering" (Meincke et al., 2025), highlighting the importance of uniqueness as well as category membership may indeed find an optimally distinctive balance (Taeuscher et al., 2021) – something that future research could confirm. On the meso level, the characteristics of the participating organizations (both on the venture and evaluator side) may also modulate the process model. For example, organizational infrastructure, learning and practices may further enable or conversely, inhibit genAI affordances from exerting their influence. Indeed, these factors may trace resonance divergence and even delegitimation (Garud et al., 2025). On the macro level, too, societal norms about the appropriateness of genAI use may have an effect. In particular, affordance visibility is affected by the current stigmatization of overt genAI use, especially, for tasks perceived as uniquely human. However, these norms may evolve over time, making genAI more accepted even for humanlike activities, such as entrepreneurial framing – and indeed many other organizational phenomena (Benk et al., 2025).

Second, GFT offers new research avenues to enrich the technological affordances literature. For succinctness, I theorized affordances as relatively stable strategic resources for both entrepreneurs and evaluators. More micro-focused investigations, however, can unearth subtle changes in how stakeholders rely on these affordances over time. For example, the use of tone repertoire may slightly differ in early stages of the process (scoping to find the right framing "voice") from later stages, when relationships are well-established (zeroing-in a single partner for optimal customization). Further, affordance visibility may also vary processually (possibly more delicately managed in the beginning of the process, and less policed in later stages). In turn, studying shifts such as these may invite further theoretical enrichment (in the above case, the engagement with marketing and business ethics scholarship on customization; Abels et al., 2025; de Bellis et al., 2019).

Third, future research should investigate the consequences of the appearance of "agentic AI," a futuristic AI model type with a robustly autonomous mode of operation, "executing complex, multistep workflows across a digital world" with minimal human input (Allouah et al., 2025; Yee et al., 2024). The appearance of such novel systems (perhaps also harbingers of General Artificial Intelligence, the holy grail of AI research) would move human-AI ensembles closer to full automation and away from human augmentation, with significant consequences (Raisch & Fomina, 2025b; Raisch & Krakowski, 2021). If founders (as well as evaluators) employ such fully agentic AI models, ghost framing will become more akin to "zombie framing": where human involvement, including affordances, strategies, resonance and legitimacy judgments, as we understand them today, are reduced to a minimum. While I argued that entrepreneurial framing will likely evade the exclusion of "humans from the loop" for many years to come, the appearance of agentic AI (as well as General Artificial Intelligence), will necessitate much further theorizing and research. At the present, this remains science fiction.

## CONCLUSION

I have brought together scholarship on entrepreneurial framing, legitimacy and sociomaterial affordances to articulate Ghost Framing Theory – a process model that explains how hybrid founder- and investor-genAI ensembles co-produce, evaluate and continually recalibrate rhetorical resonance in new venture legitimation. By theorizing five rhetorical affordances, I reveal why founder genAI strategies enhance resonance, modulated by evaluator reliance on the technology, reshaping long-standing heuristics that investors use to infer competence and authenticity. In doing so, GFT redefines the possible roles for rhetorical framing in legitimation, connects research on human-AI collaboration with cultural entrepreneurship and extends affordance theory into multi-actor scenarios where transitivity and visibility emerge as key considerations. Yet, this framework is only a first step. I invite scholars to probe its boundary conditions – such as individual skill, organizational factors and category norms – to test when ghost framing enhances resource acquisition and when it may backfire. In sum, GFT aims to spur research that treats genAI not merely as a tool but as an agentic partner whose ghost hand is already redrafting the stories launching tomorrow's ventures.

**TABLE 1**

**Summary of GenAI Rhetorical Affordances**

| GenAI Rhetorical Affordances | Definitions | Examples of GenAI Features Supporting Affordances |
|---|---|---|
| *Generativeness* | Allowing users to produce large amounts of textual (or multimodal) output at minimal input. | • Prompts generating ready-to-use content, "no matter what"<br>• "Reasoning" models' provision of detailed explanation for rationale for output<br>• Search-the-web features built into genAI apps and genAI apps built into Internet search engines |
| *Extreme combinatorics* | Allowing users to rearrange users' pre-existing inputs without the need to have a clear, humanly ordered structure for these inputs. | • Upload feature of common file formats<br>• Large context windows (size of allowable inputs such as text and files)<br>• Skill in synthetizing disparate priors into fluent posteriors |
| *Tone repertoire* | Allowing users to express themselves in a variety of rhetorical styles, matching the imagined speaker's and audience's personas. | • The range of writing styles of human communicators and content producers<br>• The voice characteristics of mimicked human speakers (e.g., NotebookLM's podcast hosts) |
| *Velocity/energy* | Allowing users to overcome human limitations of productivity and motivational energy to accomplish tasks under time pressure. | • The speed with which output is generated by genAI<br>• The non-reactance of genAI agent to repeated, arduous and large volume requests |
| *Shared substratum* | Allowing users access to our shared base of human experiences – though in a somewhat unreliable and biased manner. | • The training characteristics of genAI models, which incorporate vast swathes of the Internet<br>• Increasing size and sophistication in training of new genAI models |

**TABLE 2**

**Summary of GenAI Framing Strategies**

| Framing Strategy | Mechanism | Ensemble type | Stage | Role of GenAI Affordances |
|---|---|---|---|---|
| Robust scaffolding | Creating plausible and easy-to-digest rhetorical structures in a rapid manner about the new venture | Founder-genAI | Ghost pitching | • *Generativeness*: allows "rhetorical prototyping" with minimal effort<br>• *Velocity/energy*: allows entrepreneur to "leap-frog" over mental fatigue and problems with the venture concept, which may seem temporarily insurmountable |
| Multidimensional balancing | Resolving contradictions between different elements of the new venture's business model and its set of assumptions | Founder-genAI | Ghost pitching | • *Extreme combinatorics:* allows the simultaneous consideration of seemingly incommensurate pieces of information or representations of interest<br>• *Velocity/energy:* makes multidimensional optimization of divergent criteria possible, rapidly and through many iterations |
| Context expansion | Discovering and integrating into the pitch diverse perspectives, ways of thinking or modes of self-expression that | Founder-genAI | Ghost pitching | • *Tone repertoire:* helps stylistic matching between founders and evaluators<br>• *Shared substratum:* allows the discovery |

| Framing Strategy | Mechanism | Ensemble type | Stage | Role of GenAI Affordances |
|---|---|---|---|---|
| | are different from that of the founders (but useful in gaining resonance from evaluators) | | | of symbolic regimes unknown or unpalatable for founders |
| Rapid vetting | Processing large amounts of rhetorical pitch documents automatically based on Retrieval-Augmented Generation (RAG) | Evaluator-genAI | Ghost screening | • *Extreme combinatorics:* allows highlight multidimensional predictive criteria to enter the screening process<br>• *Velocity/energy:* facilitates initial evaluation on a scale that would be impossible for humans |
| Speed distillation | Summarizing and formulating actionable recommendations for screened-in ventures for multiple (including internal evaluator) audiences | Evaluator-genAI | Ghost screening | • *Generativeness:* allows autonomous production of reports with strategic recommendations<br>• *Shared substratum:* the complex nature of sense-making is helped by large repositories of intersubjective material in genAI's training data |
| Power feedback | Providing rhetorical responses to ventures at scale with relevant content and tone | Evaluator-genAI | Ghost screening | • *Tone repertoire:* provides the ability to generate feedback matching the founders' circumstances and framing rhetoric<br>• *Velocity/energy:* replaces the impossibly large amount of time it would take to provide |

| Framing Strategy | Mechanism | Ensemble type | Stage | Role of GenAI Affordances |
|---|---|---|---|---|
| | | | | feedback using manual labor |
| Hyperpersonalization | Following initial evaluator feedback, strategically fine-tuning subsequent framing attempts to fit underlying evaluator needs | Founder-genAI | Ghost relationship-building | • *Tone repertoire:* allows framing to match evaluator needs in both content and style (down to the word level)<br>• *Velocity/energy:* facilitates the production of large volumes of rhetorical text with an attention to detail, which is not possible by human labor |
| Enhanced responsiveness | Increasing the perceived speed of addressing evaluator requests, while promulgating a perception of a "coachable" founder persona | Founder-genAI | Ghost relationship-building | • *Shared substratum:* enables founders to reduce uncertainty based on the coherence intersubjective training data provides<br>• *Tone repertoire:* the stylistic matching of founder language to that of evaluators enhances the perception of receptiveness |
| Proactive surveillance | Anticipating and preventing likely problems that may derail startups, relying on early-stage interventions | Evaluator-genAI | Ghost relationship-building | • *Extreme combinatorics:* facilitates probing of inconsistencies and gaps in information provided by founders on an ongoing basis,<br>• *Velocity/energy:* allows coping with volume, inherent complexity and uncertainty of founder information |

| Framing Strategy | Mechanism | Ensemble type | Stage | Role of GenAI Affordances |
|---|---|---|---|---|
| Bias mitigation | Controlling many known (and unknown) evaluative decision-making biases by relying on a broader spectrum of experiences than those of the evaluator. | Evaluator-genAI | Ghost relationship-building | • *Generativeness:* allows frequent and automated communication, which is devoid of idiosyncratic biases (though may carry systematic ones inherent in genAI training)<br>• *Shared substratum:* provides access to large-scale intersubjective experience, which controls for common decision-making biases |

**FIGURE 1**

**The Ghost Framing Process**

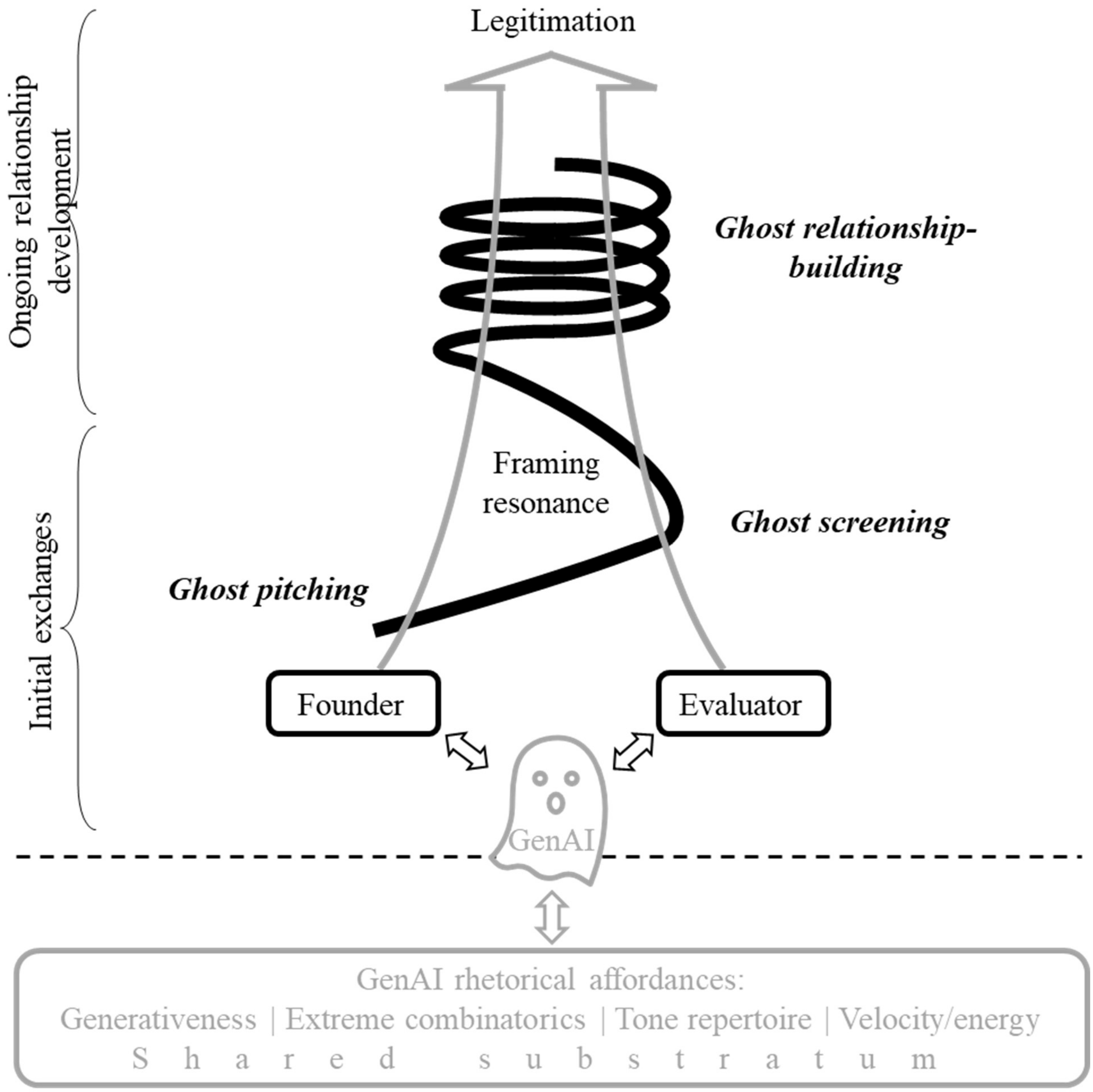

**FIGURE 2**

**Theoretical Propositions**

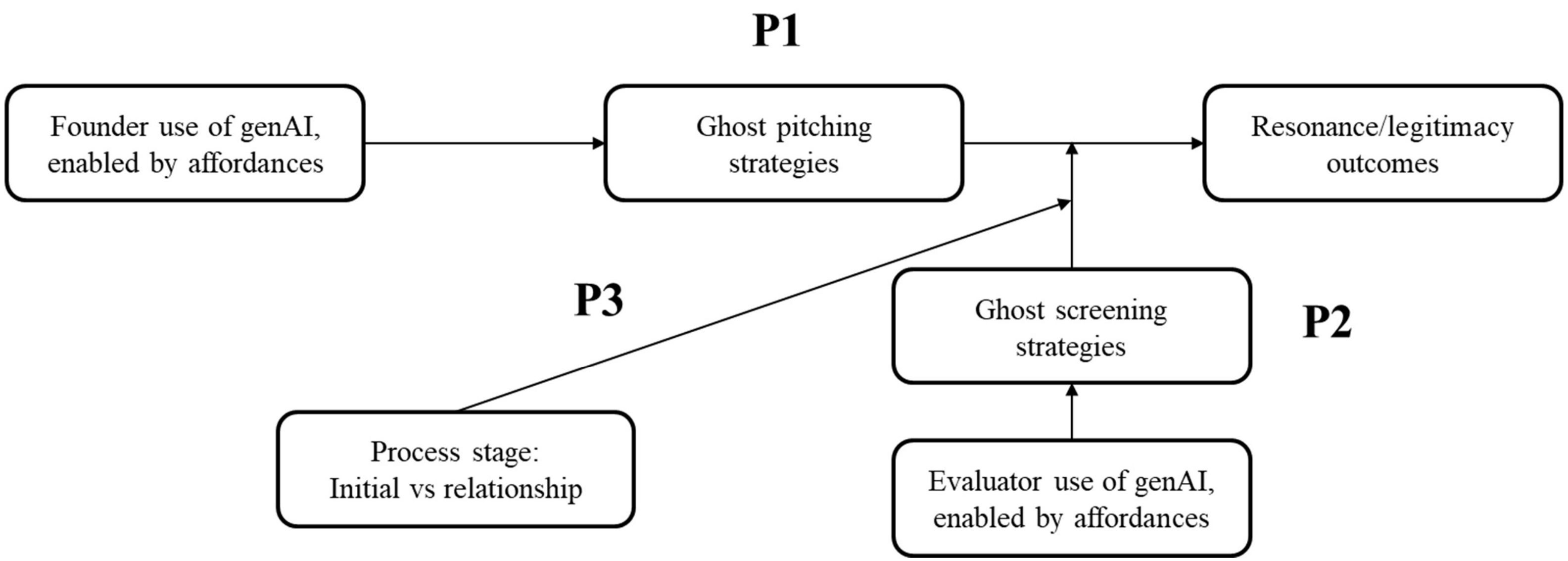